\documentclass[%
 aip,
 amsmath,amssymb,
 reprint,%
]{revtex4-1}

\usepackage{xcolor}

\usepackage{graphicx}% Include figure files
\usepackage{dcolumn}% Align table columns on decimal point
\usepackage{bm}% bold math
\usepackage[utf8]{inputenc}
\usepackage[T1]{fontenc}
\usepackage{mathptmx}
\usepackage{etoolbox}

\makeatletter
\def\@email#1#2{%
 \endgroup
 \patchcmd{\titleblock@produce}
  {\frontmatter@RRAPformat}
  {\frontmatter@RRAPformat{\produce@RRAP{*#1\href{mailto:#2}{#2}}}\frontmatter@RRAPformat}
  {}{}
}%
\makeatother
\begin{document}

\preprint{AIP/123-QED}

\title[Emergence of innovations under reputation-driven interactions]{Emergence of innovations in networked populations with reputation-driven interactions}
% Force line breaks with \\
\author{Pablo Gallarta-S\'aenz}%
\affiliation{Department of Condensed Matter Physics, University of Zaragoza, 50009 Zaragoza (Spain).}
\affiliation{GOTHAM lab, Institute for Biocomputation and Physics of Complex Systems (BIFI), University of Zaragoza, 50018 Zaragoza (Spain).}
\author{Hugo Pérez-Martínez}%
\affiliation{Department of Condensed Matter Physics, University of Zaragoza, 50009 Zaragoza (Spain).}
\affiliation{GOTHAM lab, Institute for Biocomputation and Physics of Complex Systems (BIFI), University of Zaragoza, 50018 Zaragoza (Spain).}
\author{Jes\'us G\'omez-Garde\~nes}
\email{gardenes@unizar.es}
\affiliation{Department of Condensed Matter Physics, University of Zaragoza, 50009 Zaragoza (Spain).}
\affiliation{GOTHAM lab, Institute for Biocomputation and Physics of Complex Systems (BIFI), University of Zaragoza, 50018 Zaragoza (Spain).}
\affiliation{Center for Computational Social Science, University of Kobe, 657-8501 Kobe (Japan).} 

\date{\today}% It is always \today, today,

\begin{abstract}
In this work, we analyze how reputation-based interactions influence the emergence of innovations. To do so, we make use of a dynamic model that mimics the discovery process by which, at each time step, a pair of individuals meet and merge their knowledge to eventually result in a novel technology of higher value. The way in which these pairs are brought together is found to be crucial for achieving the highest technological level. Our results show that when the influence of reputation is weak or moderate, it induces an acceleration of the discovery process with respect to the neutral case (purely random coupling). However, an excess of reputation is clearly detrimental, because it leads to an excessive concentration of knowledge in a small set of people, which prevents a diversification of the technologies discovered and, in addition, leads to societies in which a majority of individuals lack technical capabilities. 
\end{abstract}

\maketitle

\begin{quotation} 
In human societies, the concept of reputation stands as a formidable force, intricately shaping our social fabric and influencing how we engage with one another. This study delves into the profound implications of prestige in the intricate processes of human culture accumulation. We examine the common belief that interacting with knowledgeable individuals, often regarded with reputation, is a catalyst for innovation. However, our exploration reveals a nuanced reality. While preferential engagement with knowledgeable or prestigious individuals indeed expedites innovation, an excessively exclusive concentration of technological prowess among them can lead to systemic homogeneity. This homogeneity, in turn, poses a potential obstacle to technological breakthroughs, particularly those dependent on unexplored paradigms. Moreover, this concentration of expertise results in a society with diminished overall skill diversity, where a select few possess the capability to achieve higher-level innovations. Our findings emphasize the importance of a balanced interaction strategy that acknowledges the dual dynamics of fostering expertise concentration for complex innovation while concurrently nurturing a well-informed society to preserve diversity and open-mindedness, essential for exploring diverse and complementary approaches.
\end{quotation}

\section{Introduction}
\label{sec:introduction}

Human societies arise from an intricate network of interactions between distinct, autonomous human beings filled with their own motivations and desires, mostly unfathomable to the eyes of physics. However, as complex as the paradigmatic human might be, prior research has shown that the fundamental features of her conduct can be distilled into simple physical models capable of reproducing observed emergent phenomena, thus helping identify the underlying mechanisms behind human behavior~\cite{castellano2009review,jusup2022social}.

Complex systems have become an invaluable tool for the study of ever more facets of human sociality, comprising fields as diverse as sociology, psychology, economy, urban planning, or epidemiology. Some prime examples of this are models capable of reproducing observed segregation in neighborhoods based on discrimination~\cite{schelling1971segregation} or sustained cultural differences between interacting populations~\cite{axelrod1997culture}. The striking emergence of cooperation in human interactions has also been tackled from the perspective of evolutionary game theory~\cite{Szabo2007,roca2009gameTheory,perc2013gameTheory}, as well as the appearance or breaking down of coordination and consensus~\cite{baronchelli2018consensus}, and their possible relationship with prestige or indirect reciprocity \cite{xia2023reputation}. Increasing effort has been devoted to the understanding of the processes of opinion formation~\cite{deffuant2000opinion, redner2019voter} and political polarization~\cite{perez-martinez2023polarization, ojer2023polarization}, fueled by growing concerns about their negative impact on social relationships~\cite{chen2018thanksgiving} and public health~\cite{centola2011homophily, carballosa2021epidemics, fard2023epidemics}. Related to this, the spreading of behaviors, rumors and ideas have also been the object of intense research~\cite{daley1964rumours, granovetter1978threshold, gomezgardenes2016contagion, perez-martinez2022gangs}, including the propagation of news and misinformation~\cite{vicario2016misinformation, vosoughi2018fake}. All these examples show that complex systems can help shedding light over the distinct mechanisms at play in the convoluted processes of human interaction.

One of the topics that has been extensively addressed from the perspective of complex systems is the construction and propagation of human culture~\cite{boyd1985culture, derex2020evolving,pablo2022networkThinking}. The development of increasingly complex human knowledge has allowed us to survive and thrive even in the harshest environments, a skill that seems lacking in other animal species~\cite{dean2014cumulative, derex2021human, migliano2022origins} and cannot be attributed to the intelligence of single individuals and their interaction with the wild, but rather to an incremental tendency of culture accumulation, by which subsequent generations receive, evolve, and transmit a vast body of knowledge in a kind of ``ratchet effect''~\cite{tomasello1999culture, tennie2009ratcheting}. In this process, our capacity of developing innovations and the unique ability to learn from others become the keystones in the emergence of cumulative culture~\cite{boyd2011niche}. By combining existing pieces of knowledge, individuals can open up new paths of technology~\cite{dosi1982technology, youn2015invention} in a collaborative way, where interaction plays a key role driven by the sharing of different skills and perspectives that can breed innovations and a deeper understanding of the world~\cite{rogers1995innovations, tomasello1993learning, tomasello2016learning}.

Knowledge has become so important in human societies that the manifestation of excellence or expertise in some field attracts attention from others and provides a higher status~\cite{henrich2001prestige, boyd2011niche, chudek2012prestige,small2013ties}. People usually try to become closer to successful individuals in an attempt to gain benefits by acquiring either relevant skills or tangible goods~\cite{tooby1996banker}. To obtain the experts' favor, they confer deference to them creating a public conception of prestige~\cite{henrich2001prestige} and, therefore, knowledgeable individuals become salient members of the society and usual recipients of interaction~\cite{small2013ties, henrich2004demography}. However, determining the degree of knowledge of an individual can become a very difficult task, especially for newcomers or novices who may lack key information about her performance. Even so, status or deference received by experts is expected to correlate with expertise and can be easily measured, and thus looking for prestigious and widely connected individuals becomes a good strategy when lacking information about personal knowledge~\cite{henrich2001prestige, henrich2003culture, creanza2017culture}.

There are several ways by which a follower can acquire technology from such prestigious individuals, but most of them require close proximity with the models or even active interaction with them~\cite{tomasello1993learning}. An agent can obtain useful skills by observation and accurate imitation, mimicking the actions performed by their role models or their objectives (imitative learning)~\cite{henrich2001prestige,chudek2012prestige}, or by being recipients of their teaching (instructed learning)~\cite{fogarty2011teaching,tomasello2016learning}. Either the case, close relationships like friendship or kinship are needed to gain access to the desired knowledge. One way of studying how these kinds of behaviors impacted the formation of human culture is going back in time, until the early stages of humankind, and focus our attention on the interactions which took part in hunter-gatherer societies, as their lifestyle has been predominant for the longest part of human history crucially shaping our culture. These societies presented some features which allowed our ancestors to stand out over other life forms, and their success might be linked to the disposition of interacting within their groups to learn new skills~\cite{hewlett2011huntergatherer, salali2019bayaka, migliano2018speech} and the implementation of cultural transmission as a mechanism that enhances the group fitness~\cite{salali2016huntergatherer, padilla-iglesias2022musical, mesoudi2008transmission}. Nowadays, despite being present only in a few locations around the world, modern hunter-gatherer societies~\cite{migliano2017network} act as living laboratories where researchers can shed light about the origins of human culture and use them as real scenarios to put into context results from mathematical models~\cite{migliano2020cultural}.

Our aim in this study is to apply the full potential of network science to tackle the effect of prestige on the dynamic processes of cultural accumulation \cite{henrich2004demography,smolla2019structure,pablo2022networkThinking}. Following a previously developed model by Derex and Boyd~\cite{derex2016partial}, we consider innovation as a collaborative phenomenon in which two individuals have to provide their previous knowledge in order to create a new technology. When an advancement is made, it becomes known by the two agents involved in its creation, and to all those agents attached to them by any kind of link to those previously mentioned. With the aim of analyzing the effect of prestige on these processes, we consider that agents preferentially seek for what they judge as the best partners in the whole population when choosing with whom to interact in the innovation process. However, as previously stated, prestige stems from expertise, which sometimes can be difficult to estimate. Therefore, we analyze different scenarios in which \textit{(i)} agents can accurately determine the degree of knowledge of the rest of the population and \textit{(ii)} agents ignore the expertise, but know which people are the most connected, which could be considered a proxy for prestige. 

The paper is organized as follows: in Sec.~\ref{sec:model}, we introduce the innovation model as well as the chosen dynamical processes that include random and preferential interactions toward prestigious individuals. In Sec.~\ref{sec:results}, we present the main results, including the effect of preferential interactions in the innovation speed and culture spread. Finally, in Sec.~\ref{sec:conclusions}, we summarize the main findings of our work and future lines of research.

\section{The Model}
\label{sec:model}

As introduced above, our approach builds upon an adaptation of the model proposed by Derex and Boyd~\cite{derex2016partial} for the emergence and diffusion of innovations. In the original framework, the dynamic state of each individual $i$ ($i=1,...,N$) is defined by a vector $\vec{r}_i(t)$ representing cultural knowledge. These vectors have $M$ binary entries so that if an individual, say, $i$, has item $\alpha$ in her cultural repertoire at time step $t$, then the $\alpha$th entry of  $\vec{r}_i(t)$ is $\left(\vec{r}_i(t)\right)_{\alpha}=1$, while $\left(\vec{r}_i(t)\right)_{\alpha}=0$ otherwise. In this way, we consider technology as a set of defined pieces of knowledge, in line with previous studies~\cite{dosi1982technology, shennan2001demography, derex2018fragmentation}.

An essential aspect of the Derex-Boyd model is that each innovation is associated with an intrinsic value (fitness) $f_{\alpha}$ (with $\alpha=1,...,M$). Thus, considering the cultural repertoire, $\vec{r}_i(t)$, of individual $i$, we can quantify her cultural score at a given time $t$, denoted as $s_i(t)$, as
\begin{equation}
s_{i}(t)=\vec{r}_i(t)\cdot\vec{f}=\sum_{\alpha=1}^{M} \left(\vec{r}_i(t)\right)_{\alpha}\cdot f_{\alpha}\;.
\end{equation}

Equipped with these descriptors, $\vec{r}_i(t)$ and $s_i(t)$, capturing the dynamical state of each agent, we now delve into how these states evolve over time.

\subsection{Dynamical Processes}
\label{dynamics}

The agent-based simulation commences ($t=0$) with the assumption that all individuals possess identical cultural repertoires. Each member of the community is initially aware of a set of six items, $\vec{r}_i(0)=(1,1,1,1,1,1,0,\ldots,0)$ $\forall i$, each with distinct intrinsic values ($f_{1},\ldots,f_{6}$), and consequently, the same cultural score
\begin{equation} 
s_i(0)=\sum_{\alpha=1}^{6}f_{\alpha}\;.
\label{eq:InitialScore}
\end{equation}

With this initial condition, the agent-based simulations proceed as a series of iterations, each defined by the interaction of a pair of agents. At each iteration, a focal agent (say, $i$) and a target agent (say, $j$) are selected. The interaction involves the combination of the agents' knowledge to form a triad of items in the following way: one of the two agents chosen randomly contributes with two ingredients, while the other completes the triad by choosing a single item from her corresponding repertoire. Ingredients are chosen without repetition so that only 20 different triads can be formed from the initial repertoire. 

Importantly, the choice of ingredients is not random, but biased toward items of high values. In particular, the probability that item $\alpha$ is selected from the focal agent $i$'s available knowledge is
\begin{equation}
P_{i}(\alpha)=\frac{\left(\vec{r}_i(t)\right)_{\alpha}\cdot f_{\alpha}}{s_{i}(t)}\;.
\end{equation}
In this way, those items in the cultural repertoire of agent $i$ having high fitness are more frequently chosen while those less valuable are underrepresented in the innovation trials.

\begin{figure*}[t!]
\centering
\includegraphics[width=0.98\linewidth]{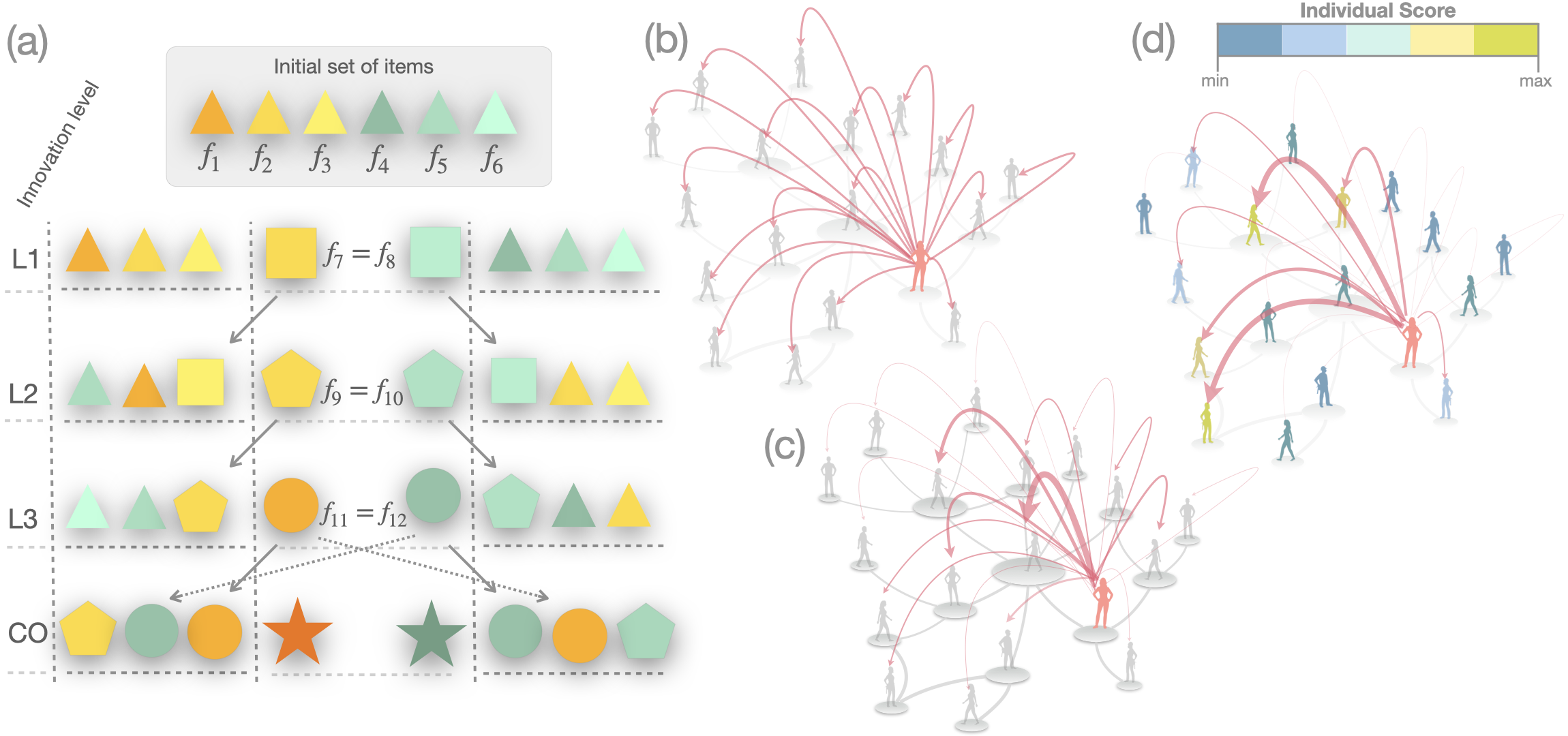}
\caption{(a) Innovation chart showing the different items of the repertoire, the specific triads of items which lead to new discoveries and the two lineages (left and right columns). The particular values for the fitness of the items used in our simulations are: $f_1=f_4=1.0$, $f_2=f_5=0.8$, $f_3=f_6=0.6$, $f_7=f_8=4.8$, $f_9=f_{10}=10.9$ and $f_{11}=f_{12}=18.8$. Note that the highest-value innovations, denoted as crossovers (CO), do not have fitness associated since they are not considered eligible items nor they count to compute $s_i(t)$. Figure and fitness adapted from Ref.\cite{derex2016partial} (b), (c) and (d) Schematic description of the target agent selection: random, degree-driven and score-driven, respectively. Agents in red represent the focal agent, while the remaining are the possible target agents. Random selection implies homogeneous probability among the whole population. Degree-driven selection is carried out according to the degree of each agent (static selection). Finally, score-driven selection identifies those agents with a higher instant score (dynamic selection) and assigns them a larger selection probability.}
\label{fig:1}
\end{figure*}

Once the triad is selected, it is tested against the innovation chart in Fig.~\ref{fig:1}(a) to determine if the combination leads to a new discovery. If successful, the new item is added to the knowledge of both agents $i$ and $j$ and their cultural score increases accordingly. If unsuccessful, nothing happens. It is crucial to note that an individual's cultural repertoire continually expands, so when a new item is discovered, the corresponding component of vector $\vec{r}_i$ changes from $0$ to $1$ and remains fixed throughout the dynamic process.

In the case of discovery, cultural transmission takes place and the acquaintances of both $i$ and $j$ also acquire the new knowledge. The transference occurs without error, incorporating the ability of high-fidelity information transmission distinctive of human beings \cite{lewis2012transmission,dean2014cumulative}. In our case, the acquaitances of each agent $i$ consist of the nearest neighbors in a social network in which each individual is associated with a node while links account for social (e.g., friendship) ties. Thus, cultural transmission occurs as information diffusion on graphs. It is important to clarify that this social network serves only for cultural transmission and not for choosing the focal and target agents who meet in the innovation process, since they are chosen from the entire population following the rules explained below in Sec.~\ref{sec:mating}.

The simulation comprises a sequence of iterations, with innovations being discovered and transmitted through social links, mimicking cultural accumulation processes in human societies. It concludes when a pair of agents reaches the highest-value innovations, described below as crossovers.

\subsection{Emergence of Incremental and Crossover Innovations}
\label{innovations}

As detailed earlier, the innovation chart in Fig.~\ref{fig:1}(a) dictates which combinations of items lead to the discovery of innovations. The chart incorporates some noteworthy properties. First, among the 20 possible combinations of the 6 initial items, only two triads produce a cultural innovation (the remaining 18 are considered non-successful). These two successful combinations lead to innovations of the first level (L1) with fitness, $f_7=f_8$, significantly larger than those of the initial items. Second, innovations beyond level L1 cannot be reached solely by combining initial items; they require the incorporation of at least one item from the previous innovation levels. For instance, to discover innovations in levels L2 and L3, a triad must be formed with the innovation discovered in L1 and L2, respectively. Additionally, the innovations in these levels have significantly higher fitness than those in previous levels, i.e., $f_7=f_8\ll f_9=f_{10}\ll f_{11}=f_{12}$.

Finally, the former evolution from the initial set of items to innovations in levels L1, L2, and L3 progressively creates two different lineages or trajectories [left and right columns of Fig.~\ref{fig:1}(a)]. These two lineages are independent since, to achieve the cultural advance in one trajectory, no innovations of the other lineage are needed. However, this independence is disrupted to discover innovations of level 4. The innovation of this last level represents a rarer leap and requires the latest two innovations from one lineage (corresponding to L2 and L3) and the last discovery of L3 in the other trajectory. With these three ingredients, the finest innovation is achieved and, due to the merging of the two lineages, this last level is labeled the crossover innovation. In this way, the model includes both the phases of ``exploitation'' or ``optimization'', the gradual change and build-up of well-defined technological trajectories, and ``exploration'' or ``innovation'', the finding of new combinations or hybridization of such trajectories, inherent to invention processes~\cite{youn2015invention, derex2021human}.

\subsection{Target agent selection mechanisms}
\label{sec:mating}

To round off the presentation of the agent-based model for the emergence and diffusion of innovations, we focus on how mating between focal ($i$) and target ($j$) agents occurs in each iteration of the simulation. The focal agent [agents in red in Figs.~\ref{fig:1}(b)-~\ref{fig:1}(d)] is randomly chosen from the set of $N$ individuals. However, for the choice of the second (target) agent, three different procedures are implemented:
\begin{itemize}
\item Random choice [Fig.~\ref{fig:1}(b)]: similar to the focal agent, the target one is randomly selected, assigning to each individual in the population a homogeneous probability,
\begin{equation}
\Pi^{R}_{j}=\frac{(1-\delta_{ij})}{N-1}\;.
\label{eq:R_prob}
\end{equation}

\item Degree-driven (DD) selection [Fig.~\ref{fig:1}(c)]: the target agent is chosen by assigning to each individual a probability that depends on their degree in the underlying social network. The probability that agent $j$ is selected as the target agent of focal agent $i$ can be written as
\begin{equation}
\Pi^{DD}_{j}=\frac{(1-\delta_{ij})k_j^{\gamma}}{\sum_{l=1}^{N}(1-\delta_{il})k^{\gamma}_l} \; ,
\label{eq:DD_prob}
\end{equation}
where $k_i$ corresponds to the degree of agent $i$. This selection rule accounts for the case in which agents are oblivious to the technological capacity of the rest of the agents, and estimate their prestige by the position in the network.

\item Score-driven (SD) selection [Fig.~\ref{fig:1}(d)]: the target agent is chosen by assigning to each individual a probability that depends on the cultural score. The probability that agent $j$ is selected as the target agent of focal agent $i$ can be written as
\begin{equation}
\Pi^{SD}_{j}=\frac{(1-\delta_{ij})s_j^{\gamma}(t)}{\sum_{l=1}^{N}(1-\delta_{il})s^{\gamma}_l(t)} \;.
\label{eq:SD_prob}
\end{equation}
\end{itemize}

\begin{figure*}[t!]
\centering
\includegraphics[width=0.97\linewidth]{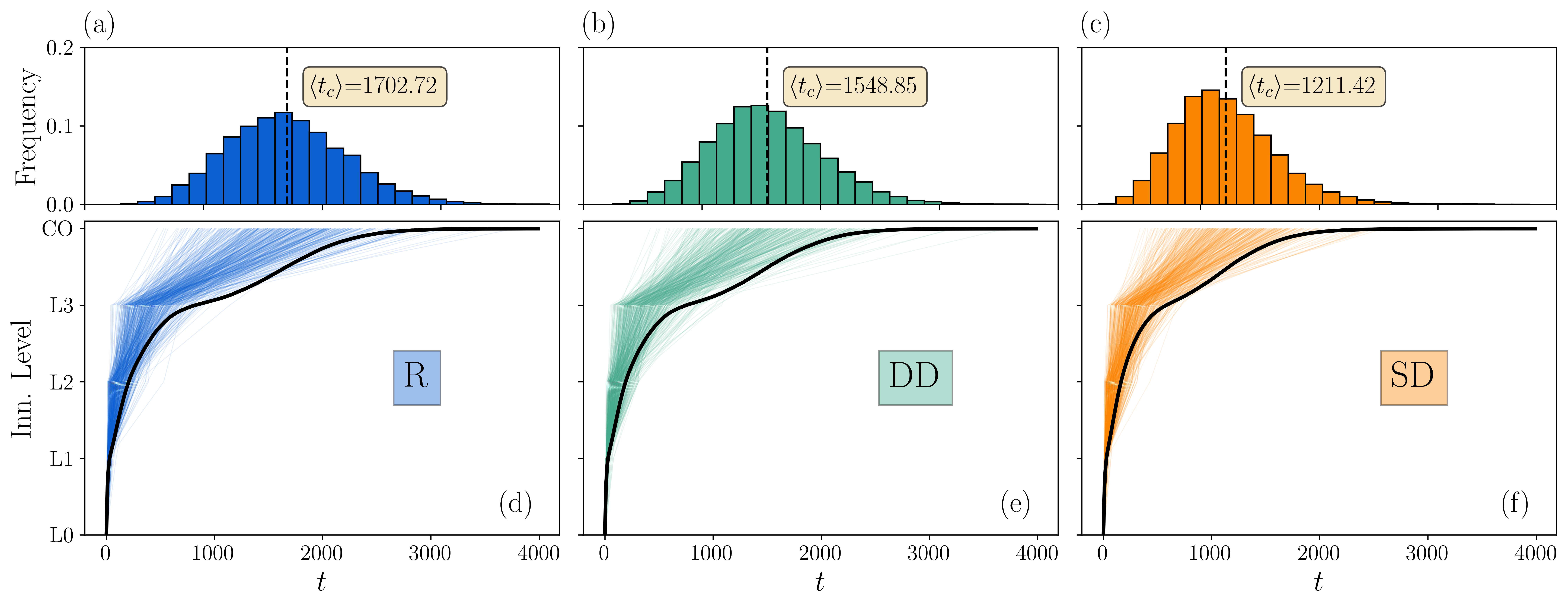}
\caption{(a), (b) and (c) Time to crossover, $t_c$, distributions obtained from $n=2\times 10^4$ independent simulations, using the three mechanisms of choosing the target agent: random (R), degree-driven (DD) and score-driven (SD), respectively. The value of driving strength to compute (\ref{eq:DD_prob}) and (\ref{eq:SD_prob}) is $\gamma=1$. (d), (e) and (f) Technological frontiers of the trajectories (colored lines) and average technological level (black line) for the three mechanisms of choosing the target agent (R, DD, and SD, respectively). Each point of every trajectory represents the time when a new technology level appears in the population. Although simulations finish when crossover is achieved, we make use of times higher than $t_c$ to compute the average technological level. For the sake of clarity, only 2.5\% of the trajectories have been represented. A decrease in the mean crossover time can be observed in the case of driven interactions with respect to the random case. The graph used throughout the simulations to transmit discoveries is an Erd\"os-R\'enyi network of $N=10^3$ nodes with average connectivity $\langle k\rangle=6$.}
\label{fig:2}
\end{figure*}

Note that the two driven mechanisms, DD and SD, differ in their static and dynamic nature, respectively. While DD favors the same set of agents from start to end of the simulation, the SD selection mechanism varies over time and starts from an homogeneous distribution since $s_i(0)$ is the same for all individuals [see Eq.~(\ref{eq:InitialScore})]. In addition, both mechanisms incorporate an exponent $\gamma$ that allows us to tune the driving strength while they both become equal to the random choice when $\gamma=0$.

\section{Results}
\label{sec:results}

We now focus on the analysis of the results obtained through extensive simulations of the agent-based model subject to different selection mechanisms, particularly those in which driving is at work.

\begin{figure}[b!]
\centering
\includegraphics[width=0.98\linewidth]{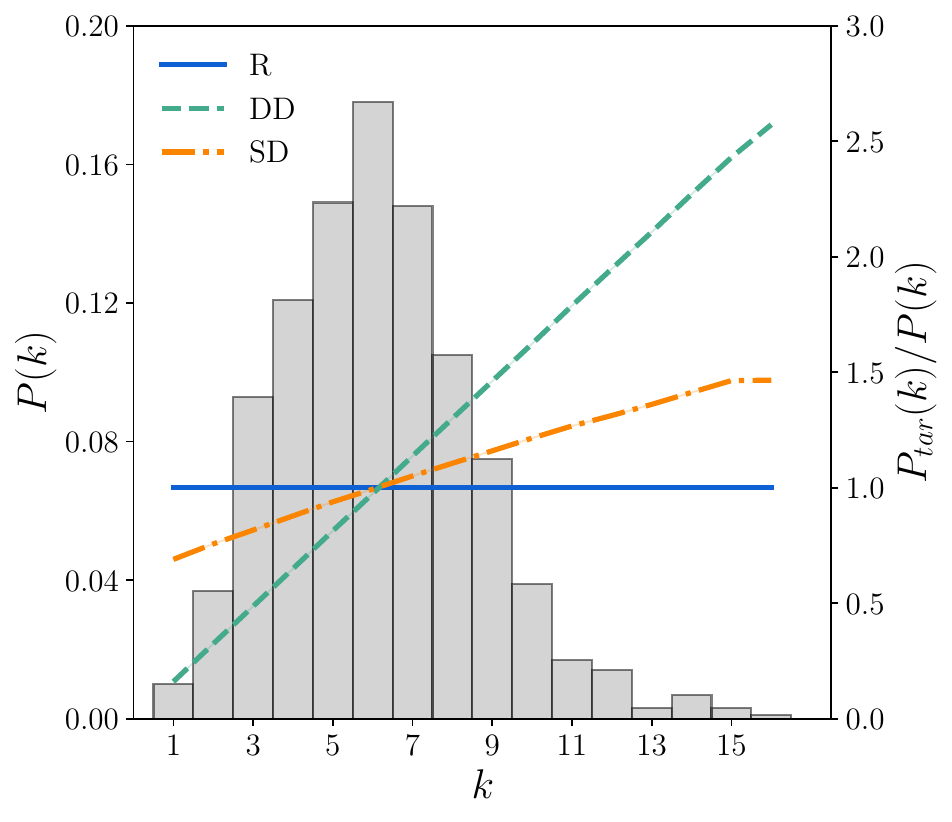}
\caption{Degree distribution $P(k)$ of the Erd\"os-R\'enyi network with $N=10^3$ nodes and average connectivity $\langle k\rangle=6$ used in the transmission of discoveries (the left axis) and probability of choosing an agent of degree $k$ as target, $P_{tar}(k)$, normalized with $P(k)$, considering the three different cases defined in the article the right axis). The value of strength is $\gamma=1$ in both driven cases. It is shown that the degree-driven strategy is much more efficient in concentrating the interaction in a small set of highly connected individuals.}
\label{fig:3}
\end{figure}

First, we analyze the linear cases ($\gamma=1$) of the driven selection mechanisms and compare these results with the random (original) case. In Fig.~\ref{fig:2}, we present the distribution of the durations required to attain the crossover innovation, denoted as $t_c$, for (a) random, (b) degree-driven, and (c) score-driven selection of target agents. These results stem from an analysis of $n=2\times 10^4$ distinct agent-based simulations on top of a random (Erd\"os-R\'enyi) graph of $N=10^3$ nodes with average connectivity $\langle k\rangle=6$ (the corresponding Poissonian degree distribution is shown in Fig.~\ref{fig:3}). The bottom graphs of Fig.~\ref{fig:2} indicate the technological frontiers of those trajectories at a given time, defined as the maximum technology level reached by any agent of the system. Trajectories culminate in the ultimate step reaching level $4$, i.e., when the crossover innovation is uncovered. From these temporal evolutions, the distributions of crossover times, $t_c$, and their corresponding averages, $\langle t_c\rangle$, it becomes clear that the SD mechanism facilitates a swifter convergence toward crossover innovations compared to the random and DD cases. The latter two exhibit comparable rates of convergence, with slight advantage for the case of DD selection.

The large differences in the crossover time suggest that the SD strategy yields a very different target selection than the other two alternatives. To confirm this hypothesis, we compute the probability of choosing an agent of degree $k$, denoted as $P_{tar}$, normalized with the degree distribution, $P(k)$, as it is shown in Fig. \ref{fig:3}. Obviously, the DD choice follows a growing linear function. However, for the SD strategy, we observe a similar growing linear function, yet with a smaller slope than in the DD scenario. This indicates that although nodes with larger degrees have higher scores than the average, their selection is not as probable as in the DD case. In this sense, it can be concluded that, while interacting with prestigious individuals is always positive, concentrating the interactions on a small set of highly connected agents might not be as beneficial as fixating in knowledgeable ones.

Now we focus on analyzing in depth the role that the strength of the driving (both at the level of degree or scores) plays on the time needed to reach the crossover. To this aim, we consider a wide range of $\gamma$ values ($\gamma\in[0,3.5]$) and for each value of $\gamma$, in a similar fashion as we did for obtaining the results shown Fig.~\ref{fig:2}, we perform $n=2\times 10^4$ distinct agent-based simulations. After averaging over the values of $t_c$ obtained in the simulations for each value of $\gamma$, we obtain the function $\langle t_c\rangle(\gamma)$.

\begin{figure}[t!]
\centering
\includegraphics[width=0.99\linewidth]{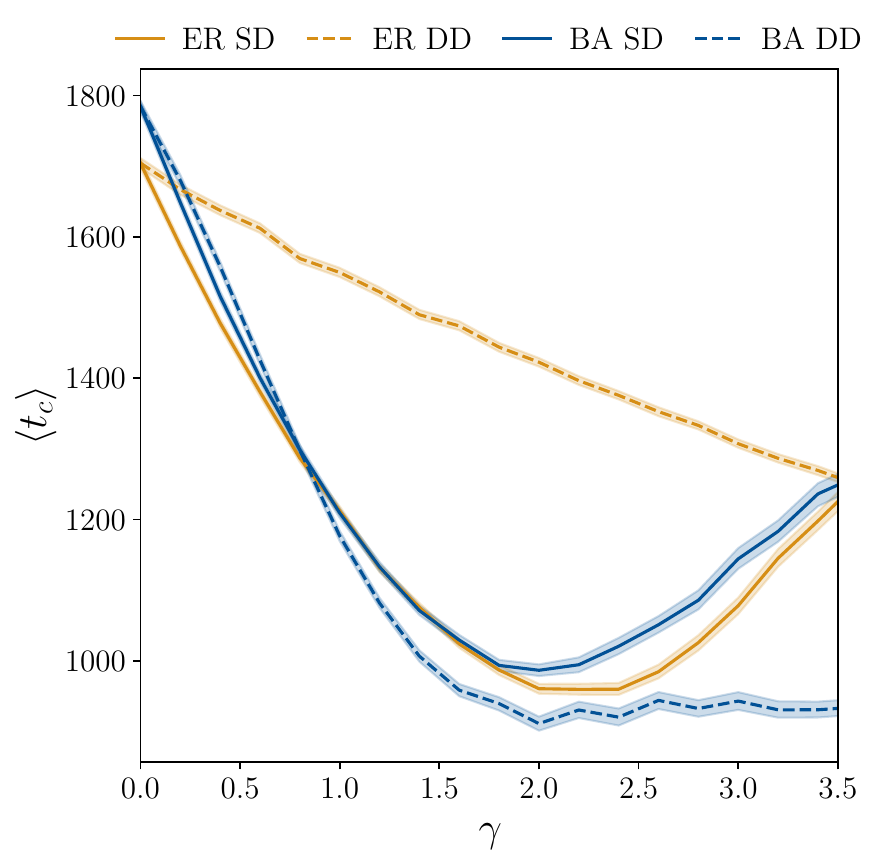}
\caption{Average time to crossover $\langle t_c \rangle$ as a function of $\gamma$,  performing $2\times10^4$ simulations for each value of $\gamma$. Results presented with 95\% confidence intervals, shown as shadowed regions. The two structures used for transmission are an Erd\"os-R\'enyi (ER) network with $N=10^3$ and $\langle k\rangle=6$ (yellow lines) and a Barabási-Albert (BA) network, with $N=10^3$ and $\langle k\rangle=6$ (blue lines). Continuous lines represent the score-driven (SD) situation, while the dashed lines correspond to the degree-driven (DD) case.}
\label{fig:4}
\end{figure}

In Fig.~\ref{fig:4}, we present the functions $\langle t_c\rangle(\gamma)$ for both the DD and SD scenarios, considering Erd\"os-R\'enyi (ER) and Barab\'asi-Albert (BA) networks as the underlying social graphs for cultural transmission. These networks have identical populations size $N$ and average degree $\langle k\rangle$. Several insights can be gleaned from these curves. First, in the case of $\gamma=0$ (where both DD and SD reduce to R), we observe that using an ER graph for cultural transmission results in a lower $t_c$ value compared to the degree-heterogeneous architecture of a BA network; this difference arising solely from the underlying network structure that guides the diffusion process of innovations. However, as $\gamma$ moderately increases ($\gamma>1$), the combination of DD target selection with a BA transmission graph emerges as the most efficient means to expedite reaching the crossover. This effect can be attributed to the scale-free nature of BA graphs, as the high heterogeneity in degrees can lead to the disproportionate concentration of interactions in a small set of individuals, where the crossover can swiftly appear.

Of particular interest is the distinctive form of the $\langle t_c\rangle(\gamma)$ functions for the SD cases. While the initial trend for all four cases suggests decreasing functions, the SD target selection in both ER and BA exhibits a minimum well beyond the linear case $\gamma=1$. This counterintuitive finding underscores that a selection of target agents strongly driven toward high score ones becomes detrimental to the discovery of crossovers in this regime. In fact, beyond the minimum, $\langle t_c\rangle$ increases with the strength of the driving $\gamma$, eventually reaching levels comparable to the DD case with ER transmission graphs when $\gamma=3.5$.

We now focus on shedding some light on the causes behind this minimum found in the function $\langle  t_c\rangle(\gamma)$ for the case of SD selection of the target agent. As explained in Sec.~\ref{innovations}, the innovation dynamics occurs in parallel following two different lineages, each being defined by their own incremental discoveries  [levels L1, L2, and L3 in Fig.~\ref{fig:1}(a)]. Thus, the original model dynamics (without any driving) includes the possibility that the system gets trapped due to a cascade of discoveries of high value in one of the lineages that prevent the advance in the other. Our hypothesis is that the SD selection mechanism, while being valuable to speed up convergence if the driving is moderate, turns out to be detrimental when strong enough by inducing biased trajectories toward one of the two lineages.

To validate this hypothesis, we have derived a macroscopic observable that allows us to quantify the degree of bias observed during a single simulation. For this purpose, let $\vec{I}_{\alpha}^{\beta}$ (with $\alpha\in{1,2,3}$ and $\beta\in{A,B}$) be a vector with $m$ components, where $m-1$ of them are equal to $0$ and only one component is equal to $1$. The nonzero component corresponds to the item of innovation level $\alpha$ in lineage $\beta$. Consequently, we can compute the instantaneous fraction of individuals having in their repertoire the innovation of level $\alpha$ in lineage $\beta$ as
\begin{equation}
n_{\alpha}^{\beta}(t)=\frac{1}{N}\sum_{i=1}^N\vec{r_i}(t)\cdot\vec{I}_{\alpha}^{\beta}.
\end{equation}
With these fractions calculated, we define the instantaneous bias as follows:
\begin{equation}
\rho(t)=\sum_{\alpha=1}^{3}f_{\alpha}\left(n_{\alpha}^{A}(t)-n_{\alpha}^{B}(t)\right),
\label{eq:inst_pol}
\end{equation}
where $f_{\alpha}$ represents the value of the two items corresponding to innovation level $\alpha$ (recall that this value is independent of the lineage). Therefore, the instantaneous bias $\rho(t)$ takes positive (negative) values when the trajectory is tilted toward lineage A (B). In addition, the greater the bias, the larger the absolute value of $\rho(t)$. 

\begin{figure*}[t!]
\centering
\includegraphics[width=0.98\linewidth]{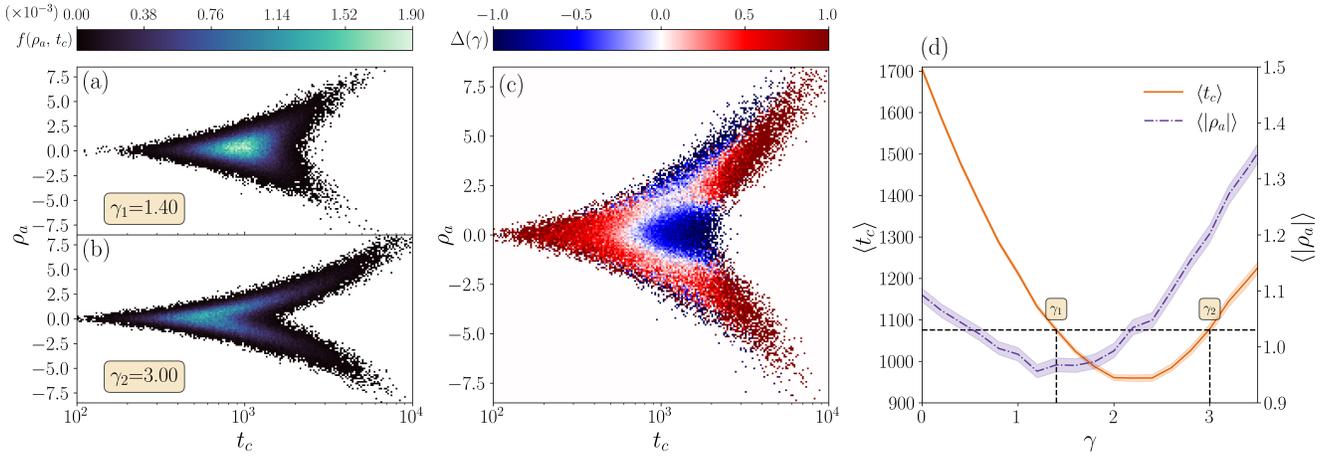}
\caption{(a) and (b) Relative frequency distribution of technological average bias $\rho_a$ and time to crossover $t_c$, $f(\rho_a,\, t_c)$. The values of $\gamma$ are selected such that their $\langle t_c\rangle$ match. In this example, $\gamma_1=1.40$ and $\gamma_2=3.00$, respectively. (c) Relative change between panels (a) and (b), defined as $\Delta(\gamma) =[f(\rho_a,\, t_c)(\gamma_2)-f(\rho_a,\, t_c)(\gamma_1)]/[f(\rho_a,\, t_c)(\gamma_2)+f(\rho_a,\, t_c)(\gamma_1)]$ for each point. (d) Average time to crossover $\langle t_c \rangle$ (the left axis) and average unsigned technological bias $\langle |\rho_a| \rangle$ (the right axis) as a function of $\gamma$. Results are shown with 95\% confidence intervals (shadowed regions). The target selection mechanism utilized is SD throughout the figure. Results are obtained from $10^5$ simulations on panels (a)-(c), and $2 \times 10^4$ simulations for each value of $\gamma$ on panel (d), using an Erdös-Rényi network with $N=10^3$ and $\langle k\rangle=6$ to transmit discoveries.}
\label{fig:5}
\end{figure*}

The former equation (\ref{eq:inst_pol}) can be used to monitor the evolution of the bias along the trajectory of a simulation until the crossover is finally reached. Therefore, it is possible to measure the extent of the technological trapping during the whole simulation by computing the average bias, $\rho_a$, i.e., the sum from $t=0$ to $t_c$ of the instant values, $\rho(t)$
\begin{equation}
\rho_a=\dfrac{\sum_{t=0}^{t_c}\rho(t)}{t_c}\;.
\label{eq:acc_pol}
\end{equation}
In Figs. ~\ref{fig:5}(a) and ~\ref{fig:5}(b) we plot the pairs of values ($\rho_a,\,t_c$) obtained for $n=10^5$ different simulations. Each panel captures the results of a set of simulations of the SD selection mechanism with ER transmission graph for two values of $\gamma$: (a) $\gamma=1.40$ and (b) $\gamma=3.00$, respectively. These two values are strategically selected being one ($\gamma=1.40$) before the minimum and the other one ($\gamma=3.00$) after it while, more importantly, for both values the average time to reach the crossover $\langle t_c\rangle$ is the same. In addition, the boxes in the two-dimensional histogram in Figs. ~\ref{fig:5}(a) and ~\ref{fig:5}(b) incorporate the relative frequency of each pair ($\rho_a,\,t_c$) (see color code).

Despite the similar values of $\langle t_c\rangle$ for both values of $\gamma$, the corresponding scatterplots show distinctly different shapes for the cloud of dots. The relative change between both situations is shown in Fig.~\ref{fig:5}(c). The most notable difference is the substantial increase in the heterogeneity of crossover times $t_c$ after the minimum, compared to the case before the minimum: most trajectories seem to reach the crossover faster, but some of them become trapped in one of the lineages (higher $|\rho_a|$), resulting in higher $t_c$. This happens because greater values of $\gamma$ correspond to societies that accumulate interactions in an increasingly small set of very knowledgeable individuals, allowing them to quickly develop the crossover. However, if this small set becomes trapped in one of the lineages (and consequently the rest of the society), then the crossover will be greatly delayed, turning the preferential interaction detrimental in some situations. Depending on which effect becomes dominant, $\langle t_c\rangle$ will increase, decrease or remain approximately the same. This is what happens in the BA DD scenario, where we have seen that a disproportionate concentration of interactions in a small set of individuals can outperform the SD strategy. In this case, the high concentration of interactions in the hubs is also able to compensate the negative effects of the bias for increasing values of $\gamma$, resulting in a nearly constant mean crossover time as a function of $\gamma$ once the minimum is reached.

For a conclusive assessment of the trapping induced by the strength of the driving in the SD selection mechanism, we compute the average unsigned bias over a set of $n$ simulations, each having an average bias of $\rho_{a}^{(i)}$ ($i=1,\ldots,n$)
\begin{equation}
\langle |\rho_a|\rangle=\dfrac{\sum_{i=1}^{n}|\rho_{a}^{(i)}|}{n}\;.
\label{eq:abs_pol}
\end{equation}
By plotting the curve $\langle |\rho_a|\rangle(\gamma)$ together with that of $\langle t_c\rangle(\gamma)$ [see Fig.~\ref{fig:5}(c)] we observe that $\langle |\rho_a|\rangle(\gamma)$ shows also a minimum that is reached for values of $\gamma$ smaller than that of $\langle t_c\rangle(\gamma)$. However, once reached the minimum, the average bias remains roughly constant for a range of $\gamma$ values and then start to increase in the region where the minimum of $\langle t_c\rangle(\gamma)$ is attained. The similar shape and correlation between the two curves indicate that, as hypothesized, the existence of a minimum for $\langle t_c\rangle(\gamma)$ is indicative of a trapping induced by the strong selection of high-scoring agents.

\begin{figure*}[t!]
\centering
\includegraphics[width=0.95\linewidth]{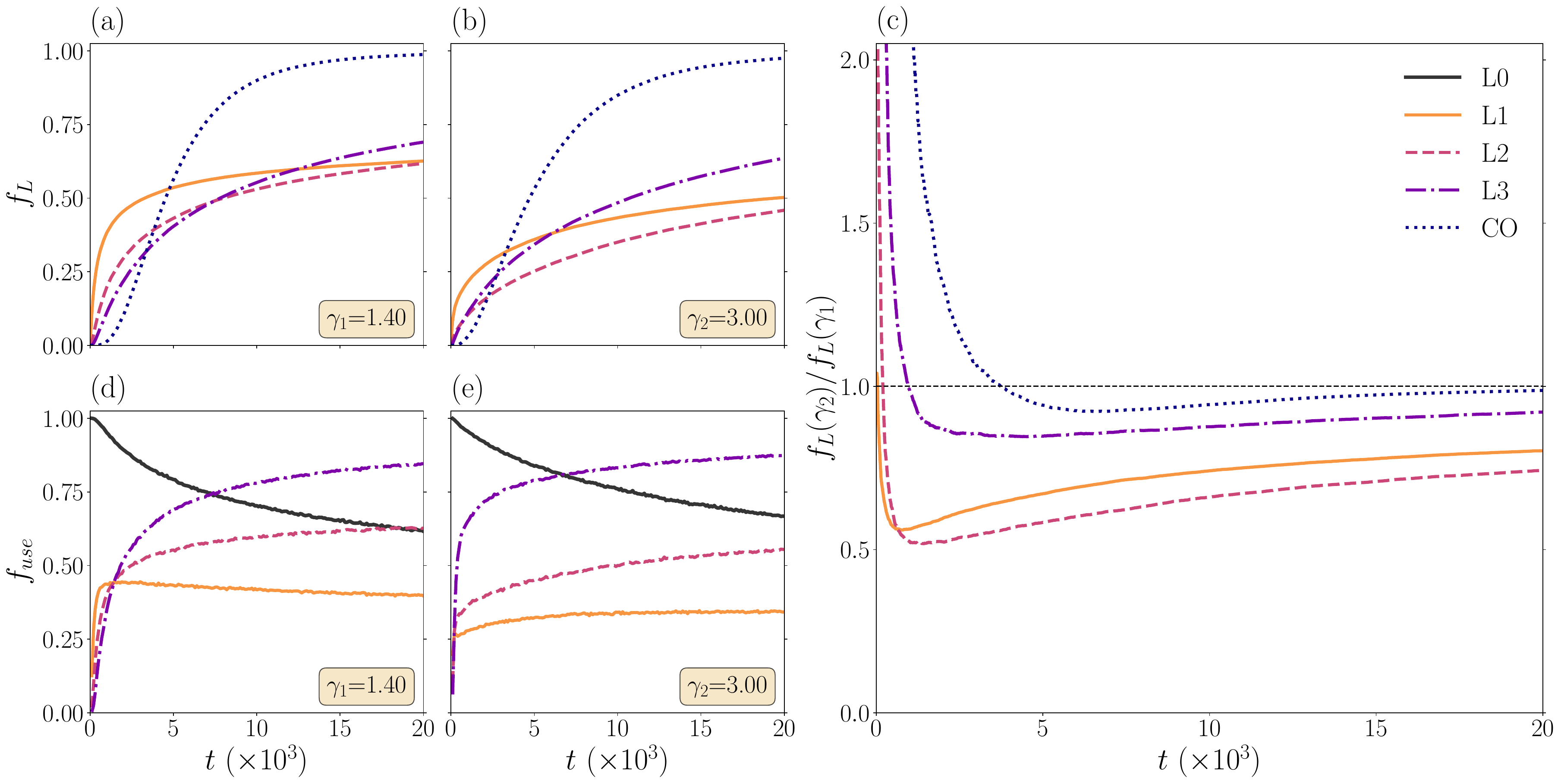}
\caption{(a) and (b) Temporal evolution of fraction of agents who have acquired an item of level L (regardless of the lineage), $f_L$. (c) Ratio between both fractions as a function of time. It can be seen that, for higher values of $\gamma$, knowledge is obtained faster initially (ratio bigger than one for the first steps), but reaches fewer agents over time (ratio smaller than one later in the simulations). Panels (d) and (e) show the frequency of use of ingredients of each level L as a function of time. Lower level innovations gradually fall into disuse. Results are shown for the two values of $\gamma$ selected in Fig.~\ref{fig:5}: (a) and (d) $\gamma_1=1.40$, (b) and (e) $\gamma_2=3.00$, respectively. Results are averaged from 500 simulations, each one concluding when $t_{max}=2\times 10^4$ is reached. For each simulation, the data point for a given time is computed by averaging the respective frequencies over the previous 100 steps. Results correspond to an Erdös-Rényi network with $N=10^3$ and $\langle k\rangle=6$, and SD selection mechanism.} Error bars are not shown because they are too small to be visible.
\label{fig:6}
\end{figure*}

The study performed above remarks on the importance of knowledgeable individuals in the dynamics, as they attract most of the interactions and quickly become bearers of the best technologies. This, however, can create huge inequalities in access to higher-level innovations of the vast majority of the population, as well as significant differences in their technological repertoires. To analyze this effect, we now perform simulations for a fixed number of steps, allowing them to continue after the crossover has appeared. To restrict ourselves to the situations previously studied in which the crossover was not present at any time during the trajectories, we assume that it can be reached and propagated, but not proposed for new triads nor counted for the agents' scores.

To address the degree of penetration of each level of technology in the population, we represent in Figs. ~\ref{fig:6} (a) and ~\ref{fig:6}(b) the fraction of agents $f_L$ that have acquired at least one technology of level L at each time step for the two values of $\gamma$ used before that correspond to roughly the same $\langle t_c\rangle$. The most remarkable result is that the crossover reaches far and wide in the whole system while the rest of innovations fall clearly short of reaching the majority of the population. This is due to the combinatorial nature of the crossover, together with the fact that more advanced innovations with higher scores are more frequently used as ingredients: it is more probable for two agents to choose the necessary ingredients, and the trapping in one of the lineages of most of the agents is not enough to prevent its apparition as long as a minority still advances in the other one. Something similar happens with the high-level technologies, as L3 can become more widespread than the lower levels, thanks to the increased probability of its ingredients being proposed.

There are also notable differences in the penetration of each level for each $\gamma$. As stated before, higher values correspond to situations in which a small set of knowledgeable individuals hoards most of the interactions and accumulate the highest levels of technology, quickly giving rise to the crossover that, however, appears in a less instructed society. On the other hand, when the interaction is more democratic, an increasingly large number of agents are able to interact and obtain intermediate technologies before the crossover is reached, smoothing the knowledge inequalities in the network. To quantify this behavior, we show in Fig.~\ref{fig:6}(c) the temporal evolution of the ratio between the fractions of the population that know each technological level for both values of $\gamma$ studied. From this plot, we observe that, after a tiny transient in which the ratio is larger than $1$ (resulting from the faster trajectories that appear for higher values of $\gamma$), agents evolving according to the larger $\gamma$ value become systematically more ignorant of higher-level technologies than in the lower $\gamma$ situation. This behavior remains well after the crossover has been reached for the first time, concluding that, indeed, the concentration of knowledge leads to faster innovation at the cost of generating less informed societies.

The fact that different pieces of knowledge are not shared by the majority of the population coincides with what happens in real-world hunter-gatherer communities, in which single individuals rarely know and use all the technologies accessible to their societies~\cite{salali2016huntergatherer}. These results also show that it is not necessary to know every step that leads to a new technology in order to use it fluently, as observed, for example, in the adoption of western medicine in such hunter-gatherer societies when it is readily available~\cite{reyes-garcia2013loss, salali2020weirding}, or the use of GPS by the Inuits instead of traditional means of path finding~\cite{aporta2005gps}.

Lastly, in Figs.~\ref{fig:6} (d) and ~\ref{fig:6}(e) we represent, for each level, the fraction of interactions in which at least one ingredient of said level has been proposed as a function of time. It becomes clear that higher level ingredients are preferentially chosen to the detriment of the lower level ones, which appear less frequently. Even though the model is too simple to render a technology obsolete, this result qualitatively reproduces the loss of traditional technologies when more efficient ones are available~\cite{aporta2005gps, reyes-garcia2013loss, salali2020weirding}.

\section{Conclusions}
\label{sec:conclusions}

Prestige is a key asset in human societies, capable of shaping how we interact with each other. In this work, we have explored its effect on the processes of human culture accumulation, considering the fact that knowledgeable individuals are usually revered and mirrored, and finding that seeking for expertise in the innovation process might not always be the best option. We have found that preferentially interacting with knowledgeable or prestigious individuals greatly accelerates innovation, but if this preference is strong enough so that technology becomes an exclusive asset of such individuals, then the increased homogeneity of the system could force innovation into a dead end, delaying technological breakthroughs that rely on the undeveloped paradigms. Moreover, this concentrating effect results in a less skilled society, in which fewer individuals are able to attain higher level innovations. 

These results indicate that the optimal interaction strategy must consider multiple effects: on the one hand, there is a beneficial outcome in the accumulation of culture in a relatively small set of individuals capable of developing increasingly complex innovations. On the other hand, keeping a well-informed society is key to provide enough plurality and open-mindedness needed to explore multiple complementary approaches.

There are several limitations to our approach. First, we have adopted an innovation model with a narrow scope, as few technologies are considered and collaborative interaction is needed in order to discover them. In reality, the attainable set of technologies is gargantuan~\cite{youn2015invention} and the number of individuals involved in the innovation processes that generate them can range from a single person to whole corporations~\cite{rogers1995innovations}. Nevertheless, the aforementioned model encompasses some key aspects of the innovation process, namely the phases of exploitation of established technological trajectories, and innovation based on the combination of those trajectories~\cite{youn2015invention, derex2021human}. Related to this, the available technological set is so reduced that we are not able to observe clearly the processes of technological loss in the face of higher level technologies that occur in real-world societies~\cite{aporta2005gps, reyes-garcia2013loss, salali2020weirding}. However, even with this limited number of technologies, a notable decrease in the use of outdated ingredients can be observed, that hints at this process being reproducible with our simple assumptions.

Second, we have only addressed the effect of prestige, neglecting other network features that are known to impact the processes of culture accumulation. For example, we have not explored the influence of connectedness over the phenomena observed in our work, while previous studies show that highly connected networks seem to decrease its cultural diversity and innovation capacities~\cite{derex2016partial, derex2018fragmentation}. Moreover, demography and population size have also been identified as key modulators of the technological advancement, as larger societies might be able to host a wider set of innovations avoiding an excessive technological homogeneity~\cite{shennan2001demography, derex2020evolving}, while smaller ones might lead to the quick demise of their culture~\cite{henrich2004demography}. Some recent studies have also shown that hunter-gatherer societies possess a multilevel structure comprised of family, camps, and whole regions that can provide unique diffusion networks that enhance the processes of culture accumulation~\cite{migliano2017network,migliano2020cultural}. In addition, mobility, migration and changing network topologies that emerge in such communities can promote innovation while maintaining relatively heterogeneous cultures~\cite{derex2018fragmentation,smolla2019structure,garg2021foraging}. Lastly, we have assumed homogeneity in the interaction strategies, while in reality multiple choices coexist and contribute to the wider technological advancement~\cite{mesoudi2008experimental, mesoudi2008agentbased,mesoudi2008successful}.

In summary, our study constitutes an important step in the field, addressing the need of considering prestige and reputation in the way humans interact. Although great amounts of work remain to be done in order to unveil the relevant characteristics that crucially shape the processes of culture accumulation and technological innovations, our results provide a solid ground for future developments that take into account multiple mechanisms already proven to influence how our societies have become what they are today.

\begin{acknowledgments}
P.G.S., H.P.M. and J.G.G. acknowledge financial support from the Departamento de Industria e Innovaci\'on del Gobierno de Arag\'on y Fondo Social Europeo (FENOL group grant E36-23R), and from Ministerio de Ciencia e Innovaci\'on through project PID2020-113582GB-I00/AEI/10.13039/501100011033. P.G.S. acknowledges financial support from the European Union-NextGenerationEU and Servicio P\'ublico de Empleo Estatal through Programa Investigo 024-67. H.P.M. acknowledges financial support from Gobierno de Arag\'on through a doctoral fellowship.
\end{acknowledgments}

\section*{Data Availability Statement}
Data sharing is not applicable to this article as no new data were created or analyzed in this study.

\section*{References}
\bibliography{aipsamp}% Produces the bibliography via BibTeX.

\end{document}